\documentstyle[12pt]{article}
\def\beq{\begin{equation}}
\def\eeq{\end{equation}}
\def\bea{\begin{eqnarray}}
\def\eea{\end{eqnarray}}
\def\bq{\begin{quote}}
\def\eq{\end{quote}}

\def\bbz{ Z \kern-8.9pt Z}
\def\gappeq{\mathrel{ \rlap{\raise.5ex\hbox{$>$}}
{\lower.5ex\hbox{$\sim$}}}}
\def\lappeq{\mathrel{ \rlap{\raise.5ex\hbox{$<$}}
{\lower.5ex\hbox{$\sim$}}}}
\begin{document}

\begin{flushright}
{CERN-TH.6285/91}\\
\end{flushright}
\vspace*{5mm}
\begin{center}
{\bf ANALYSIS OF ALL DIMENSIONFUL PARAMETERS RELEVANT IN
GRAVITATIONAL DRESSING OF CONFORMAL THEORIES} \\
\vspace*{1cm}
H. Dorn
\footnote{Permanent address: IEP, Humboldt Universit\"at, Berlin}\\
Theory Division, CERN, CH-1211 Geneva 23, Switzerland
 \\
\vspace*{1cm}
H.J. Otto \\
Institut f\"ur Elementarteilchenphysik, Humboldt-Universit\"at,\\
Invalidenstr.110,~D-O-1040 Berlin, Germany \\
\vspace*{2cm}
{\bf ABSTRACT} \\
\end{center}
\vspace*{5mm}
Starting from a covariant and background independent definition of
normal ordered vertex operators we give an alternative derivation of
the KPZ relation between conformal dimensions and their gravitational
dressed partners. With our method we are able to study for arbitrary
genus the dependence of N-point functions on {\em all} dimensionful
parameters. Implications for the interpretation of gravitational dressed
dimensions are discussed.
\vspace*{8cm}
\begin{flushleft}
CERN-TH.6285/91 \\
November 1991 \\
\end{flushleft}
\thispagestyle{empty}
\newpage
\setcounter{page}{1}
\pagestyle{plain}
\section {Introduction}
If a two dimensional conformal theory, specified by the central charge $c$,
the conformal dimensions $\Delta_{i}^{(0)}$ of its primary fields and the
operator product expansion coefficients, is coupled to 2D gravity the fields
get new scaling dimensions $\Delta_{i}$ which obey the KPZ relation
\cite{r1,r2,r3}
\beq
\Delta_{i} - \Delta_{i}^{(0)} = \frac{(\sqrt{1-c}-\sqrt{25-c})^{2}}{24} \Delta
_{i}(1-\Delta_{i}).
\label{e1}
\eeq
The original derivation \cite{r1} is based on the use of the light cone gauge
and $\Delta_{i}$ is related to scaling in the coordinate $z^{-}$. In the more
familiar conformal gauge any relation between scaling in geometrical length
and scaling in coordinates is lost after quantization of the metrics. Here the
dimensions $\Delta_{i}$ are defined \cite{r2,r3} via the area dependence
of N-point functions of the so called gravitational dressed primary fields,
calculated for fixed area $A$ and vanishing Liouville mass (cosmological
 constant) $m^{2}=0$
\beq
\frac{\langle \prod _{i}( V_{i}(z_{i}))_{dressed}\rangle _{A,m^{2}=0}}{Z
_{A,m^{2}=0}} \propto A^{\sum_{i}(1-\Delta_{i})}.
\label{e2}
\eeq
$Z_{A,m^{2}=0}$ is the corresponding partition function. In refs. \cite{r2,r3}
the system is treated as a conformal theory in a background metrics
$ \hat{g}_{ab}(z)$ and the characteristic parameters as the string
susceptibility
and the exponents of the dressing factors are fixed by the requirement of
background independence of both $Z_{A,m^{2}=0}$ and the integrated primary
fields.
All manipulations are performed with the formal unregularized functional
integral.

The two point function for instance behaves itself by construction as a
function of its coordinates like $\vert z_{1}-z_{2}\vert ^{-4}$ and as a
function of $A$ like $A^{2-2\Delta}$. Since (at least if the fields are
represented by scalar vertex operators) the function under discussion has
naive 2D dimension zero, there must be a further dimensionful parameter
involved. This is of course a renormalization scale $\mu $ originating from
the suppression of self-contractions. The two point function then must look
like $(\mu \vert z_{1}-z_{2}\vert )^{-4}(\mu ^{2}A)^{2-2\Delta}$. In a
previous
paper \cite{r4} we have given an alternative derivation of the KPZ relation
based on the use of an a priori background independent and general covariant
definition of normal ordered vertex operators. This procedure contains $\mu $
from the very beginning. Furthermore, we found an additional dimensional
quantity $\hat{R}_{0}$ to be relevant. It parametrizes the integration
constant in the
Liouville action. The analysis was restricted to genus zero and vanishing
cosmological constant $m^{2}=0$. Using the technique of formal continuation
in the central charge, introduced in \cite{r5} and proved to be correct in
\cite{r6}, the aim of the present paper is a complete discussion of the
dependence of N-point functions on {\em all} dimensionful parameters for
arbitrary genus of the 2D surface. The parameters are: $l$ denoting the
length scale in coordinate space, $\hat{R}_{0},\mu$ and $m^{2}$ ($A$ is the
Laplace transformed variable to $m^{2}$).
%
%
\section{Analysis for genus $h\geq 2$}
We consider correlation functions of vertex operators
\beq
V_{k}(z)= e^{ikX(z)}
\label{e3}
\eeq
in a matter theory $(X^{\mu}, \mu=1, ... ,d)$ described by the action
\beq
S=\frac{1}{8\pi}\int d^{2}z\sqrt{g(z)}~\Big (g^{mn}(z)\partial _{m}X^{\mu}(z)
\partial _{n}X_{\mu}(z)+iR(z)P^{\mu}X_{\mu}(z)\Big ),
\label{e4}
\eeq
where $R$ is the curvature scalar derived from the metrics $g_{ab}$.
The background charges $P^{\mu }$ have been introduced to describe both
noncritical strings $(P=0, d\neq 26)$ and minimal conformal theories
$(P\neq 0, d=1)$. The central charge is
\beq
c= d-3P^{2}.
\label{e5}
\eeq
After performing the $X^{\mu}$ and ghost integrations we arrive at
\bea
\langle \prod _{j=1}^{N} V_{k_{j}}(z_{j})\rangle &=& \delta \Big(\sum_{j}k_{j}-
P(1-h)\Big)
\int d\mu (\tau )
\nonumber\\
&\times &\int {\cal D} \sigma e^{-(26-d)S_{L}[e^{\sigma}g_{ab}^{(\tau )}]}
\nonumber\\ & \times &exp\Big[-\frac{1}{2}\sum_{i,j}k_{i}k_{j}G(z_{i},z_{j}
\vert e^{\sigma}g_{ab}^{(\tau)})\Big].
\label{e6}
\eea
$d\mu(\tau)$ denotes the moduli integration measure, $g_{ab}^{(\tau)}$ is a
metrics completely determined by the moduli up to
diffeomorphisms. $S_{L}[g_{ab}]$ is the Liouville action.
Its change for $g_{ab}\rightarrow e^{\sigma}
g_{ab}$ is given by the integrated conformal anomaly
\bea
S_{L}[\sigma \vert g] & \equiv & S_{L}[e^{\sigma}g]-S_{L}[g]
\nonumber \\
& = & \frac{1}{48\pi}\int d^{2}z\sqrt{g}~\Big (\frac{1}{2}g^{mn}\partial _{m}
\sigma \partial _{n}\sigma +R\sigma +m^{2}(e^{\sigma}-1)\Big ).
\label{e7}
\eea
$G$ is the Arakelov Green function \cite{r7} which obeys
\bea
G(z_{i},z_{j}\vert e^{\sigma}g) =G(z_{i},z_{j}\vert g)-\frac{\sigma (z_{i})
+\sigma (z_{j})}{2(1-h)}+\frac {6}{(1-h)^{2}}S_{L}^{0}[\sigma \vert g]
\label{e8}
\eea
where $ S_{L}^{0}[\sigma \vert g] $ is given by (\ref{e7}) with $m^{2}=0$.

The Liouville field $\sigma $ is not a scalar in the present parametrization
since $g_{ab}^{(\tau )}$ depends on the moduli only. Therefore, it is more
convenient to choose an arbitrary reference $\hat{\sigma}$ and to parametrize
the metrics with a scalar $\sigma $ by
\beq
g_{ab}(z)=e^{\sigma (z)+\hat{\sigma}(z)}g_{ab}^{(\tau )}(z)\equiv e^{\sigma}
\hat{g}_{ab}.
\label{e9}
\eeq
The integration measure ${\cal D} \sigma $ is not translation invariant
\cite{r8}.
Using \cite{r4},\cite{r9}
\beq
{\cal D} (\sigma +\hat{\sigma})= D_{\hat{g}}\sigma e^{S_L[\sigma \vert
\hat{g}]}
\label{e10}
\eeq
with translation invariant $D_{\hat{g}}\sigma $ we get from (\ref{e6})
\bea
\langle \prod_{j=1}^{N}V_{k_{j}}(z_{j})\rangle &=& \delta \Big(\sum_{j}k_{j}
-P(1-h)\Big) \int d\mu (\tau ) e^{-(26-d)S_{L}[\hat{g}]}
\nonumber\\
&\times &\int D_{\hat{g}}\sigma ~e^{-(25-d)S_{L}[\sigma \vert \hat{g}]}~
exp\Big[-\frac{1}{2}\sum_{i,j}k_{i}k_{j}G(z_{i},z_{j}\vert e^{\sigma}\hat{g})
\Big].
\label{e11}
\eea
For the discussion of the ultraviolet regularization we split
\beq
G(z_{i},z_{j}\vert g)=-log(M^{2}\vert z_{i}-z_{j}\vert ^{2}) +
G_{M}(z_{i},z_{j}
\vert g).
\label{e12}
\eeq
M is an auxiliary scale. It will be related below in a convenient way to the
parameters mentioned in the introduction. $G_{M}$, defined by (\ref{e12})
, is regular at $z_{i}=z_{j}$ and obeys the same scaling relation (\ref{e8})
 as G itself.

We now regularize the logarithm as
\beq
log(M^{2}\vert z_{i}-z_{j}\vert ^{2})\longrightarrow log\Big( M^{2}[\vert
z_{i}-z_{j}\vert ^{2}+\epsilon ^{2}/ \mu ^{2}]\Big)
\label{e13}
\eeq
with dimensionless $\epsilon $ and RG-mass scale $\mu $. Furthermore, the
normal product version of $V_{k}(z)$ is defined \cite{r4} by multiplication
with a general covariant factor whose single purpose is the cancellation of
the $\epsilon $ singularities after quantization of both $X^{\mu }$ and
$\sigma $
\beq
:V_{k}(z):~  = V_{k}(z)\Big(\epsilon ^{2}e^{\sigma }\sqrt{\hat{g} }\Big)
^{a(k)-1}e^{\sigma }\sqrt{\hat{g} }.
\label{e14}
\eeq
The last factor $ e^{\sigma }\sqrt{\hat{g}}=\sqrt{g} $ is added since after
quantization of the metrics only scalar densities make sense. With the
technique
%
%
%
%
used below the Liouville interaction term $m^{2}e^{\sigma }$ appears as an
operator insertion evaluated with $S_{L}^{0}$. To cancel also the ultraviolet
singularities of this operator we finally define as our regularized starting
point
\bea
\langle \prod _{j=1}^{N} : V_{k_{j}}(z_{j}) : \rangle &=&\delta \Big(\sum_{j}k_
{j}-P(1-h)\Big)\int d\mu (\tau )e^{-(26-d)S_{L}[\hat{g}]}
e^{\frac{m^{2}Q^{2}\hat{A}}{16\pi }} \nonumber\\
&\times &\prod _{j}\Big(\Big(\frac{M}{\mu}
\Big)^{k_{j}^{2}}\epsilon ^{k_{j}^{2}+2(a_{j}-1)}(\hat{g}(z_{j}))
^{\frac{a_{j}}{2}}\Big)
\prod_{i\neq j}(M\vert z_{i}-z_{j}\vert )^{k_{i}k_{j}}
\nonumber\\
&\times &\int D_{\hat{g}}\sigma ~e^
{-3Q^{2}(S_{L}^{0}[\sigma \vert \hat{g}]+
\frac{m^{2}}{48\pi }\epsilon ^{2(a_{0}-1)}\int d^{2}z\hat{g}
^{\frac{a_{0}}{2}}e^{a_{0}\sigma })}
\nonumber\\
&\times &\prod_{j} e^{a_{j}\sigma (z_{j})}~
exp\Big[-\frac{1}{2}\sum_{i,j}k_{i}k_{j}G_{M}(z_{i},z_{j}\vert
e^{\sigma}\hat{g})\Big] ,
\label{e15}
\eea
with $\hat{A}$ being the area corresponding to $\hat{g}$ ,
\beq
Q^{2} = \frac{25-d}{3}  ,
\label{e16}
\eeq
and
\beq
a_{i} = a(k_{i})  ,~~  a_{0} = a(0)   .
\label{e17}
\eeq

Using the scaling relation (\ref{e8}) for $G_{M} $ and taking into account
the $\delta $ -function constraint, the $\sigma $ -functional integral in
(\ref{e15}) becomes
\bea
I&=& exp\Big[-\frac{1}{2}\sum_{i,j}k_{i}k_{j}G_{M}(z_{i},z_{j}\vert
\hat{g})\Big]\int D_{\hat{g}}\sigma \prod _{j=1}^{N}
e^{(a_{j}+\frac{Pk_{j}}{2})\sigma (z_{j})} \nonumber\\
&\times &e^{-3(Q^{2}+P^{2})S_{L}^{0}[\sigma \vert \hat{g}]}~exp\Big[-
\frac{m^{2}Q^{2}}{16\pi }\epsilon ^{2(a_{0}-1)}\int d^{2}z\hat{g}
^{\frac{a_{0}}{2}}e^{a_{0}\sigma }\Big] .
\label{e18}
\eea
Now we perform the $\sigma $  integration \`{a} la ref.\cite{r5} with the
zero mode defined appropriately for the use of the Arakelov Green function
($\sigma _{0}=(8\pi (1-h))^{-1}\int d^{2}z\sqrt{\hat{g}}\hat{R}\sigma$).
For abbreviation we introduce
\beq
b_{i} = a_{i} + \frac{Pk_{i}}{2}
\label{e19}
\eeq
and
\beq
s = \frac{\frac{Q^{2}}{2}(1-h)-\sum_{i} a_{i}}{a_{0}} =
\frac{\frac{Q^{2}+P^{2}}{2}(1-h)-\sum_{i} b_{i}}{a_{0}} .
\label{e20}
\eeq
As usual we start with positive integer s and continue analytically afterwards.
{}From now on capital indices $I,J,...$ will be understood as
running between 1 and $s$, while $i,j,...$ refer to the vertex insertion points
$1,2,...,N$ as before.
 \bea
I &=&\vert a_{0}\vert ^{-1}\Gamma (-s)\Big(\frac{Q^{2}m^{2}}{16\pi }\Big)^{s}
\epsilon ^{2s(a_{0}-1)}e^{S_{L}[\hat{g}]} \nonumber\\
&\times &\Big (\frac{M\epsilon }{\mu }\Big)
^{-4\frac{\sum _{i}b_{i}^{2}}{Q^{2}+P^{2}}
-4\frac{sa_{0}^{2}}{Q^{2}+P^{2}}}
\prod_{i\neq j} (M\vert z_{i}-z_{j}\vert )
^{-4\frac{b_{i}b_{j}}{Q^{2}+P^{2}}}\nonumber\\
&\times &\int \prod_{J}\Big(d^{2}w_{J}\hat{g}(w_{J})^{\frac{a_{0}}{2}}\Big)
\prod_{I\neq J}(M\vert w_{I}-w_{J}\vert )
^{-4\frac{a_{0}^{2}}{Q^{2}+P^{2}}}
\prod_{i,J}(M\vert z_{i}-w_{J}\vert )^{-8\frac{a_{0}b_{i}}{Q^{2}+P^{2}}}
\nonumber\\
&\times&exp\Big[\sum_{i,j}(-\frac{1}{2}k_{i}k_{j}
+\frac{2}{Q^{2}+P^{2}}b_{i}b_{j})
G_{M}(z_{i},z_{j}\vert
\hat{g})\nonumber\\
&&~~~~~~+~~\frac{2}{Q^{2}+P^{2}}\Big(\sum_{I,J}a_{0}^{2}
G_{M}(w_{I},w_{J}\vert \hat{g})
+2\sum_{i,J}a_{0}b_{i}G_{M}(z_{i},w_{J}\vert \hat{g})\Big)\Big].
\label{e21}
\eea
Putting the integral $I$ back into (\ref{e15}) we find cancellation of
$\epsilon$ for
\beq
k_{i}^{2}+2(a_{i}-1)-4\frac{b_{i}^{2}}
{Q^{2}+P^{2}}
=0
\label{e22}
\eeq
which means
\beq
b_{i}=\frac{Q^{2}+P^{2}}{4}\Big(1\pm \sqrt {1+\frac{8}{Q^{2}+P^{2}}
(\Delta_{0}(k_{i})-1)}~\Big)
\label{e23}
\eeq
\beq
a_{0}=\frac{Q^{2}+P^{2}}{4}\Big(1-\sqrt{1-\frac{8}{Q^{2}+P^{2}}}~\Big) .
\label{e24}
\eeq
$\Delta_{0}(k)$ is the conformal dimension of the original $V_{k}(z)$
\beq
\Delta_{0}(k)=\frac{k(k-P)}{2}    .
\label{e25}
\eeq
As usual the sign ambiguity for $a_{0}$ has been resolved
by correspondence
with the quasiclassical limit \cite{r3} or by
requiring microscopic nature
of $e^{a_{0}\sigma } $ \cite{r10} . For comparison with other treatments we
note that our Liouville field is normalized just as it appears
in the conformal
anomaly of the $X$ integration. To compare e.g. with ref. \cite{r3} one
has to rescale $ \sigma \rightarrow -2\sigma /\sqrt{Q^{2}+P^{2}} $ .
Altogether, using (\ref{e22}) we get from (\ref{e15})
\bea
\langle \prod_{j}:V_{k_{j}}(z_{j}):\rangle &=&
\delta \Big(\sum_{j}k_{j}-
P(1-h)\Big)\int d\mu (\tau )e^{-3Q^{2}S_{L}^{0}[\hat{g}]}
\vert a_{0}\vert ^{-1}\Gamma (-s)
\nonumber\\
&\times &\Big (\frac{Q^{2}m^{2}}{16\pi }\Big)^{s}
\prod_{j}(\hat{g}(z_{j}))^{\frac{a_{j}}{2}}
\Big(\frac{M}{\mu }\Big)^{-\sum_{i}2(a_{i}-1)-2s(a_{0}-1)}
\nonumber\\
&\times &\prod_{i\neq j}(M\vert z_{i}-z_{j}\vert )
^{k_{i}k_{j}-4\frac{b_{i}b_{j}}{Q^{2}+P^{2}}}
\int \prod_{J}\Big(d^{2}w_{J}\hat{g}(w_{J})^{\frac{a_{0}}{2}}\Big)
\nonumber\\
&\times &\prod_{I\neq J}(M\vert w_{I}-w_{J}\vert )
^{\frac{-4a_{0}^{2}}{Q^{2}+P^{2}}}
\prod_{i,J}(M\vert z_{i}-w_{J}\vert )^{\frac{-8a_{0}b_{i}}{Q^{2}+P^{2}}}
J_{M}( z,w \vert \hat{g}) .
\label{e26}
\eea
The definition of $J_{M}$ is obvious from (\ref{e21}), it is the product of all
factors containing $G_{M}$ .

By construction there is no dependence on M and $\hat{\sigma}$. As a
consistency
check one can verify this for eq.(\ref{e26}) explictely by using the $\delta
$ - constraint, (\ref{e20},\ref{e22}). During this exercise one finds
\beq
J_{\lambda M}(z,w \vert \hat{g})=\lambda ^{Q^{2}(1-h)^{2}}
J_{M}(z,w \vert \hat{g})
\label{e27}
\eeq
\bea
J_{M}(z,w \vert e^{\varphi }\hat{g})=
J_{M}(z,w \vert \hat{g}) e^{3Q^{2}S_{L}^{0}[\varphi \vert \hat{g}]}
\prod_{j}e^{-a_{j}\varphi (z_{j})}\prod_{J}e^{-a_{0}\varphi (w_{J})}.
\label{e28}
\eea
Thanks to $\hat{\sigma }$- independence we now can choose $\hat{\sigma }$
to describe a constant curvature metrics. Let $\hat{R}$ denote the
corresponding curvature scalar. Then
\beq
S_{L}^{0}[\hat{g}] = S_{L}^{0}(\hat{R},\tau _{i})
\label{e29}
\eeq
is a usual function of $\hat{R}$ and $\tau_{i}$. Specializing (\ref{e7}) to
this situation we find
\beq
\frac{d}{d\hat{R}}S_{L}^{0}(\hat{R},\tau _{i}) = -\frac{1-h}{6\hat{R}}  .
\label{e30}
\eeq
This implies
\beq
S_{L}^{0}(\hat{R},\tau _{i})=S_{L}^{0}(\hat{R}/\hat{R}_{0})=
-\frac{1-h}{6}\log (\hat{R}/\hat{R}_{0})
\label{e31}
\eeq
with an integration constant $\propto \log \hat{R}_{0}$ which can be taken
$\tau _{i}$- independent. The moduli occur in the general case, of course,
\beq
S_{L}^{0}[e^{\sigma }\hat{g}]=-\frac{1-h}{6}\log (\hat{R}/\hat{R}_{0})
+\frac{1}{48\pi }\int d^{2}z\sqrt{\hat{g}}~\Big (\frac{1}{2}\hat{g}^{mn}
\partial_{m}\sigma \partial_{n} \sigma +\hat{R}\sigma \Big )
\label{e32}
\eeq
via the boundaries of the $z$-integration in a Fuchsian or Schottky type
parametrization of the surface.

Now we recognize a $\hat{R}_{0}$-dependence of (\ref{e26}) via
$S_{L}^{0}[\hat{g}]$. As before there is no dependence on $\hat{R}$ and $M$.
Therefore, we are free to put
\beq
\hat{R} = M^{2} = \hat{R}_{0} .
\label{e33}
\eeq
{}From (\ref{e27}) and (\ref{e28}) we get with (\ref{e20}),(\ref{e22})
\beq
J_{(\lambda \hat{R}_{0})^{\frac{1}{2}}}(z,w \vert \lambda ^{-1}
\hat{g}_{0})=\lambda ^{\frac{Q^{2}(1-h)^{2}}{2}}
J_{\hat{R}_{0}^{\frac{1}{2}}}(z,w \vert \hat{g}_{0}) .
\label{e34}
\eeq
Here $\hat{g}_{0}$ denotes the metric yielding $\hat{R}_{0}$ .

Now we are ready to study the scaling behaviour of (\ref{e26}). Concerning
the coordinate dependence we are interested in uniform scaling, i.e.
\beq
z_{i} =  l\cdot \xi _{i} , ~~ w_{I} = l\cdot \eta _{I}
\label{e35}
\eeq
with dimensionless $\xi _{i},\eta _{I}$. $J$ is dimensionless and depends on
two dimensionful parameters : $\hat{R}_{0}$ and $l$. Therefore, we get from
(\ref{e34})
\beq
J_{\hat{R}_{0}^{\frac{1}{2}}}(l\xi ,l\eta \vert
\hat{g}_{0})=(\hat{R}_{0}l^{2})^{\frac{Q^{2}(1-h)^{2}}{2}}
f(\xi ,\eta ,\tau ) .
\label{e36}
\eeq
A last necessary information concerns $\hat{g}_{0}$ . We use the
uniformization theorem to represent our surface as the upper half plane
divided by some subgroup of the isometry group of the constant curvature
metrics \cite{r11}
\beq
ds^{2} = \frac{-2}{\hat{R}_{0}(Im z)^{2}}dzd\bar{z} .
\label{e37}
\eeq
This implies
\beq
\Big(\hat{g}_{0}(z_{i})\Big)^{\frac{1}{2}}
=(\hat{R}_{0}l^{2})^{-1}
\Big(\hat{g}_{0}(\xi _{i})\Big)^{\frac{1}{2}} .
\label{e38}
\eeq
We use eqs. (33),(35),(36),(38), again the $\delta $-constraint and (20),(22)
and take into account that the scaling in coordinate space also affects the
boundaries but does not change the dimensionless moduli. We then get finally
\bea
\langle \prod_{j}:V_{k_{j}}(z_{j}):\rangle &=&
\Big[\Big(\frac{\mu ^{2}}{m^{2}}\Big)^{2-\gamma _{0}}\Big(\frac{\hat{R}_{0}}
{\mu ^{2}}\Big)^{\frac{25-d}{6}}\Big]^{h-1}
\Big(\frac{m}{\mu }\Big)^{2\sum_{j}\Delta _{j}}(ml)^{-2N}
F_{h,N}(\xi )
\label{e39}
\eea
with
\bea
F_{h,N}(\xi )&=&\delta \Big (\sum_{j}k_{j}-P(1-h)\Big )
\vert a_{0}\vert ^{-1}\Gamma (-s)\Big(\frac{Q^{2}}{16\pi }\Big)^{s}
\prod_{j}(\hat{g}_{0}(\xi_{j}))^{\frac{a_{j}}{2}} \nonumber\\
&\times &\prod_{i\neq j}\vert \xi _{i}-\xi _{j}\vert
^{k_{i}k_{j}-4\frac{b_{i}b_{j}}{Q^{2}+P^{2}}}
\int d\mu (\tau )
\int \prod_{J}\Big(d^{2}\eta _{J}\hat{g}_{0}(\eta _{J})^{\frac{a_{0}}{2}}\Big)
\nonumber\\
&\times &\prod_{I\neq J}\vert \eta _{I}-\eta _{J}\vert
^{-4\frac{a_{0}^{2}}{Q^{2}+P^{2}}}
\prod_{i,J}\vert \xi _{i}-\eta _{J}\vert ^{-8\frac{a_{0}b_{i}}{Q^{2}+P^{2}}}
f(\xi ,\eta ,\tau ) .
\label{e40}
\eea
The dimensions $\Delta _{i}$ are defined by
\beq
\frac{b_{i}}{a_{0}} = 1-\Delta _{i}
\label{e41}
\eeq
and $\gamma _{0}$ denotes the string susceptibility \cite{r2,r3}
$$
\gamma _{0} = 2-\frac{Q^{2}+P^{2}}{2a_{0}}=2+\frac{1}{12}
\Big(c-25-\sqrt{(25-c)(1-c)}\Big) .
$$
%
%
\section{Analysis for $h=0,1$}
The sphere fits into the above analysis up to eqs. (37,38). However, it is
straightforward to determine $J$ explicitely
\beq
J_{\hat{R}_{0}^{\frac{1}{2}}}(z,w \vert \hat{g}_{0})=
e^{\frac{Q^{2}}{2}(-1+\log 8)}
\prod_{j}e^{-a_{j}\hat{\sigma }_{0} (z_{j})}
\prod_{J}e^{-a_{0}\hat{\sigma }_{0} (w_{J})}.
\label{e42}
\eeq
Then $\hat{\sigma }_{0}$ cancels and (39) remains correct for h=0, too.
(The parameter $M$ of ref.\cite{r4} is related to $\hat{R}_{0}$ by
$\hat{R}_{0}=8M^{2}e^{-1}$. )

A little bit more thought is needed for the torus $h=1.$ In this case we cannot
use the Arakelov Green function and are forced
to exploit the more standard Green function
with the zero mode defined by weighting with $\hat{g}^{\frac{1}{2}}$ instead of
$\hat{g}^{\frac{1}{2}}\hat{R}$.
Then in the scaling relation replacing (8) there appears the integral
$ \int d^{2}z G(z_{i},z\vert \hat{g})e^{\sigma }\sqrt{\hat{g}} $ instead of the
local $\sigma(z_{i})$ \cite{r11,r12}. But fortunately, just for $h=1$ we always
have momentum conservation $\sum_{i}k_{i}=0$ and that unpleasant nonlocal
quantity drops out in the $X^{\mu }$-integration. As there is no constantly
curved torus there is no $\hat{R}_{0}$-dependence.
For the $\sigma $- integration
we can refer to flat $\hat{g}$ and use the standard expression of G by Theta
functions. After all we reproduce the dimensional factors in (39) with $h=1$.
%
%
\section{Discussion}
The scaling behaviour in $m,\mu ,l,\hat{R}_{0}$ can be directly read off from
our final result (39),(40) if $F_{h,N}(\xi )$ is finite and nonzero.
While in general one expects this to be true ( for a computation of
$F_{0,N}$ see \cite{r5,r6}), there are certain important exceptions. The first
obvious one is the case of integer $s\geq 0$. An example for such
a situation is the genus one partition function $(s=0)$. The singularity is due
to the divergent $\sigma $-zero mode integration used in going from eq.(18)
to eq.(21). For noninteger $s>0$ the analytic continuation from $s<0$ yields
finite
results, but for integer $s\geq 0$ we are just sitting on the poles of the
$\Gamma $-function. The $\sigma _{0}$-integral is divergent at $s\geq 0$ for
$\sigma_{0}\rightarrow -\infty $. This is an ultraviolet problem in geometrical
length. Therefore, our treatment of the UV-problem was not quite complete.
It cancels only divergencies due to selfcontractions of vertex operators built
from
$X^{\mu }$ and the {\em nonzero} mode of $\sigma $. We do not know at
present how to relate a cutoff in $\sigma _{0}$ \cite{r14,r15} to our
$\epsilon $ in a covariant and background independent way, since the splitting
in $\sigma =\sigma_{0}+\tilde{\sigma }$ always refers to a background
$\hat{g}$.

Fortunately, one can proceed by an indirect argument, based on observations in
\cite{r10,r16} . Since the area operator $\int d^{2}z:V_{0}:$ gives a
contribution (-1) to $s$ one finds that for an arbitrary $N$ point function
$$
\Big( \frac{d}{dm^{2}}\Big)^{[s]+1}\langle \prod _{j}:V_{k_{j}}(z_{j}):\rangle
$$
has a total $s$-value $<0$ corresponding to a convergent $\sigma
_{0}$-integration.
Applying to this quantity the now unquestioned formulas (39),(40) we reproduce
the latter after $[s]+1$ integrations for the original $s\geq 0,$ but with two
modifications. There are additional nonuniversal (i.e. with renormalization
scheme dependent coefficients) integer powers $(m^{2}/\mu ^{2})^{i}$;\ i=
0,1,...,[s]\  due to [s]+1 free integration constants. For integer s$\geq 0$
in the leading universal term $(m^{2}/\mu ^{2})^{s}$ is replaced by
$(m^{2}/\mu ^{2})^{s}log(m^{2}/\mu ^{2})$.

A further potential source of divergent $F_{h,N}$ is the moduli integration.
Since the moduli are dimensionless a corresponding cutoff does not influence
the balance of dimensionful scales in our scheme. The relation to other
statements in the literature \cite{r10,r16} still has to be clarified.

A large amount of activity has been devoted to the $c=1$ case. Here $F_{h,N}$
is zero (for an explicit discussion for $h=0$ see \cite{r5,r17}). This is due
to the fact
that at $c=1$ there has to be made a new choice for the cosmological constant
operator \cite{r10,r15}.

After this digression concerning the validity and some modifications of
(39),(40) we now turn to the discussion of the scaling properties themselves.
Formulas are always understood up to logarithmic corrections and nonuniversal
power terms. With $N=0$ eq.(39) yields the scaling property of the partition
function $Z_{h}$
\beq
Z_{h} \propto \Big [ (\frac {\mu ^{2}}{m^{2}})^{2-\gamma _{0}}(\frac {\hat
{R}_{0}
}{\mu ^{2}})^{\frac {25-d}{6}} \Big ]^{h-1}.
\label{e43}
\eeq
As mentioned in the introduction, the area is the only geometrical quantity
left after $\sigma $-integration. As familiar from \cite {r2,r3} we can divide
the $\sigma $-integration in (18) into two steps
\bea
I&=&e^{-\frac {1}{2}\sum _{i,j}k_{i}k_{j}G}\int _{0}^{\infty }dA ~e^{-m^{2}A}
\int D_{\hat {g}}\sigma ~\delta \Big (\frac {Q^{2}}{16\pi }\epsilon
^{2(a_{0}-1)}
\int d^{2}z\hat{g}^{\frac {a_{0}}{2}}e^{a_{0}\sigma }-A \Big )\nonumber \\
&\times & \prod _{j}e^{b_{j}\sigma (z_{j})}e^{-3(P^{2}+Q^{2})S_{L}^{0}}.
\label{e44}
\eea
Introduced in this manner $A$ is nothing more than the Laplace transformed
variable of $m^{2}$ and every scaling $(m^{2})^{y} $ translates into $A^{-1
-y}$. Besides this technical area variable A there is of course the expectation
value of the area $\bar{A}$
\beq
\bar{A}=\frac{\int d^{2}z \langle :V_{0}(z):\rangle }{Z_{h}}\propto m^{-2}.
\label{e45}
\eeq

The normalized N-point functions
\beq
\frac{\langle \prod _{j}:V_{k_{j}}(z_{j}):\rangle}{Z_{h}} \propto
\Big(\frac{m}{\mu }\Big)^{2\sum_{j}\Delta _{j}}(ml)^{-2N}
\label{e46}
\eeq
are independent on
$\hat{R}_{0}$.
Since $\hat{R}_{0}$ parametrizes a field independent constant in $S_{L}$
this was clear from the beginning. However, $\hat{R}_{0}$ remains present
in the sum over all genera of the integrated N-point function relevant for
string
applications.
\beq
\frac{\sum_{h}\langle \prod _{j}(\int d^{2}z_{j}:V_{k_{j}}(z_{j}):)\rangle}
{\sum _{h}Z_{h}}=\Big(\frac{m}{\mu }\Big)^{2\sum_{j}\Delta _{j}}(m)^{-2N}
\frac{\sum_{h}g_{eff}^{h}H_{h,N}}{\sum_{h}g_{eff}^{h}H_{h,0}}
\label{e47}
\eeq
with $H_{h,N}=\int \prod _{i}d^{2}\xi _{i}F_{h,N}(\xi )$ and
\beq
g_{eff}=g\Big(\frac{\mu ^{2}}{m^{2}}\Big)^{2-\gamma _{0}}
\Big(\frac{\hat{R}_{0}}{\mu ^{2}}\Big)^{\frac{25-d}{6}}.
\label{e48}
\eeq
$g$ is the string coupling. The structure of $g_{eff}$ resembles that of
the effective coupling in matrix models \cite{r13}.

A remarkable fact about (48) is the following. The contribution of $m^{2}$
is governed by the central charge $c$, that of $\hat{R}_{0}$ by the dimension
$d$ of the string target space and that of $\mu $ by both $c$ and $d$.

A last remark concerns the relation to our previous paper \cite{r4} where the
$m^{2}=0$ case was treated. Requiring finite results for $m^{2}\rightarrow 0$
one has to ensure vanishing exponents of $m^{2}$. This reintroduces the second
$\delta $-constraint produced otherwise by the $\sigma $-zero mode integration
in the massless case. Then,to get the scaling powers $\Delta_{i}$,scaling in
$\hat{R}_{0}$ is crucial ("$\lambda $-scaling").

\vspace*{1cm}
\noindent
{\bf Acknowledgements}

H.D. thanks the CERN Theoretical
Division for kind hospitality and financial
support as well as L. Alvarez-Gaum\'{e} and D. L\"{u}st for discussions.

\end{document}